# Magnetic and transport properties of iron-platinum arsenide $Ca_{10}(Pt_{4-\delta}As_8)(Fe_{2-x}Pt_xAs_2)_5$ single crystal

Qing-Ping Ding,[1,2] Yuji Tsuchiya,[1] Shyam Mohan,[1] Toshihiro Taen,[1] Yasuyuki Nakajima,[1,2] Tsuyoshi Tamegai[1,2]

[1]*Department of Applied Physics, The University of Tokyo, 7-3-1 Hongo, Bunkyo-ku, Tokyo 113-8656, Japan*

[2]*JST, Transformative Research-Project on Iron Pnictides (TRIP),*

*7-3-1 Hongo, Bunkyo-ku, Tokyo 113-8656, Japan*

We report superconducting properties of single crystalline $Ca_{10}(Pt_{4-\delta}As_8)(Fe_{2-x}Pt_xAs_2)_5$ by X-ray diffraction, magnetization, resistivity, and magneto-optical imaging measurements. The magnetization measurements reveal fish-tail hysteresis loop and relatively high critical current density $J_c \sim 0.8 \times 10^5$ A/cm$^2$ at low temperatures. The exponential temperature dependence of $J_c$, which arises from nonlinear effective flux-creep activation energy, has been observed. Upper critical field determined by resistive transition shows a relatively large anisotropy. The magneto-optical images reveal homogenous current flow within the crystal.

PACS number(s): 74.25.Ha, 74.25.Sv, 74.62.Bf

## I. INTRODUCTION

A series of iron-based superconductors have been synthesized after the discovery of superconductivity in oxypnictide $LaFeAsO_{1-x}F_x$ (1111) with a critical temperature $T_c \sim 26$ K.[1-18] Layers of Fe tetrahedrally coordinated by As or chalcogenides are common features in these materials, which is similar to $CuO_2$ layer in high-$T_c$ cuprates. In iron-based superconductors, blocking layers with alkali, alkali-earth, rare-earth oxides, alkali-earth fluorides, and complex metal oxides are alternatively stacked with Fe-As layers.[1-9,11-18] The ionic chemical bonds which appear within the blocking layers make them insulating. Recently, intermetallic Pt-As layers was introduced as new blocking layers since Pt is very flexible and can form arsenides with different coordinations.[19-23] Two structures of iron-platinum arsenides with composition $Ca_{10}(Pt_3As_8)(Fe_{2-x}Pt_xAs_2)_5$ (10-3-8) and $Ca_{10}(Pt_4As_8)(Fe_{2-x}Pt_xAs_2)_5$ (10-4-8) with $T_c$ of ~13 K and above 30 K, respectively, have been identified.[19-22] Although 'high-quality' single crystals have been reported by several groups, the homogeneity of the superconducting states has not yet been investigated. Upper critical field ($H_{c2}$) is an important parameter in superconductors. Most iron-based superconductors show relatively weak anisotropy of $H_{c2}$. In particular, $H_{c2}$ in Fe(Te,Se) is almost isotropic at low temperatures.[24] In $Ba(Fe,Co)_2As_2$, the anisotropy is also very small.[25,26] This small anisotropy is a typical difference between most iron-based superconductors and cuprate superconductors. Bearing a different structure from other iron-pnictides, the degree of the anisotropy in iron-platinum arsenides is an interesting topic. Critical current density is another eye-catching and important parameter of superconductors. The ability to carry an electrical current without dissipation enables their use in devices such as high power transmission lines, high field magnets, compact high quality filters, etc. The exploration of higher critical current density not only gives the answer needed for practical applications, but also shed light on the fundamental properties of the superconducting state itself. Above mentioned curiosities constitute the motivations of our present study.

In this paper, we report on synthesis and characterizations of high-quality $Ca_{10}(Pt_{4-\delta}As_8)(Fe_{2-x}Pt_xAs_2)_5$ single crystals. The characterization through X-ray diffraction, EDX, magnetization, and resistivity measurements are discussed. We also adopted magneto-optical (MO) imaging technique to check the homogeneity of the prepared crystal.

## II. EXPERIMENTS

Ca chips (Rare Metallic, 99.5%), As pieces (Fukuzawa Electric, 99.99999%), Pt chips (Furuya Metal, 99.95%), and FeAs powder were used as starting materials to prepare $Ca_{10}(Pt_{4-\delta}As_8)(Fe_{2-x}Pt_xAs_2)_5$ single crystals. FeAs was prepared by placing stoichiometric amounts of As pieces (Fukuzawa Electric, 99.99999%) and Fe powder (Kojundo Chemical Laboratory, 99%) in an evacuated quartz tube and reacting them at 1065 °C for 10 h after heating at 700 °C for 6 h. The mixture of 2 g with a ratio of Ca: Fe: Pt: As = 15 : 16 : 8 : 30 were loaded into an alumina crucible (inner diameter 12 mm) and sealed in an evacuated quartz tube (inner diameter 18 mm). The whole assembly was heated up to 1000 °C for 72 hours after heating at 700 °C for 5 h, followed by cooling down to 800 °C at a rate of 3 °C/h. Then the furnace was switched off and cooled to room temperature naturally.

The phase identification of the sample was carried out by means of X-ray diffraction (M18XHF, MAC Science) with Cu-$K\alpha$ radiation generated at 40 kV and 350 mA. The chemical composition of the crystal was confirmed by EDX (S-4300, Hitachi High-Technologies equipped with EMAX x-act, HORIBA). Bulk magnetization is measured by a superconducting quantum interference device (SQUID) magnetometer (MPMS-5XL, Quantum Design). Resistivity measurements were performed in the sample chamber of a SQUID magnetometer by the four-probe method with silver paste for electrical contacts. MO imaging was employed for local magnetic characterization. A Bi-substituted iron-garnet indicator film is placed in direct contact with the sample, and the whole assembly is attached to the cold finger of a He-flow cryostat (Microstat-HR, Oxford Instruments). MO images were acquired by using a cooled CCD camera with 12-bit resolution (ORCA-ER, Hamamatsu). To enhance the visibility of the local magnetic induction and eliminate the signals from the impurity phases and scratches in the garnet film, a differential imaging technique is adopted.[27,28]

### III. RESULTS AND DISCUSSIONS

Figure 1(a) shows the X-ray pattern of as-prepared single crystal. Only 00$l$ peaks are observed, and the $c$-axis lattice constant is calculated as 1.0343 nm, which is a little smaller than that of Refs. 20-22. The chemical composition of the crystal determined by EDX analyses is $Ca_{10}(Pt_{4-\delta}As_8)(Fe_{1.96}Pt_{0.04}As_2)_5$ with δ=0.49 based on the ratio of each element.

Figure 1(b) shows temperature dependence of the resistivity of $Ca_{10}(Pt_{4-\delta}As_8)(Fe_{2-x}Pt_xAs_2)_5$ single crystal. The resistivity at room temperature is relatively low with 200 μΩ cm, and shows metallic behavior in the normal state. A sharp drop in resistivity was observed starting from 30.3 K, which indicates the onset of superconductivity. The zero resistivity occurs at 28.7 K and the transition width is 1.6 K. The residual resistivity ratio (RRR) ρ(300 K)/ρ($T_c^{onset}$) is 2.15. Realtively large ρ($T_c^{onset}$) and small RRR may reflect the presence of Pt ions in FeAs layers as suggested in ref. 19-22.

Temperature dependences of zero-field-cooled (ZFC) and field-cooled (FC) magnetization at 5 Oe of $Ca_{10}(Pt_{4-\delta}As_8)(Fe_{2-x}Pt_xAs_2)_5$ single crystal are shown in Figure 2(a). The sample shows an onset of diamagnetism at $T_c$~33 K. Figure 2(b) shows the magnetization as a function of field at several temperatures ranging from 2 to 25 K.

From the magnetization hysteresis loops, we can evaluate global critical current density $J_c$ for a single crystal using the Bean's model with the assumption of field–independent $J_c$. According to Bean model,[29] $J_c$ [A/cm$^2$] is given by

$$J_c = 20 \frac{\Delta M}{a(1 - a/3b)}. \quad (1)$$

where $\Delta M$ is $M_{down} - M_{up}$, $M_{up}$ and $M_{down}$ are the magnetization when sweeping field up and down, respectively, $a$ and $b$ are sample widths ($a<b$). Figure 3(a) shows the field dependences of caculated $J_c$ from the data shown in Fig. 2(b) using Eq. (1) and sample dimensions $a$~800 μm and $b$~1190 μm.

A pronounced nonmonotonic field dependence of $J_c$ with a broad maximum, fish-tail effect, is clearly observed at temperatures higher than 10 K as shown in Fig. 2(b). The same effect is believed to exist below 10 K with much higher peak field above 50 kOe. The fish-tail effect is more clearly seen in the field dependence of $J_c$ as shown in Fig. 3(a). First of all, we define the fish-tail effect in the present context. This name is invented in the study of high temperature cuprate superconductors, which often shows a broad peak in the magnetic field dependence of the critical current density. Such a broad peak in $J_c$-$H$ is ubiquitous among cuprate superconductors,[30,31] except for very anisotropic $Bi_2Sr_2CaCu_2O_{8+y}$ compounds.[32] It is also widely observed in iron-based superconductors except for some isovalently doped system such as $BaFe_2(As,P)_2$,[33] although some other report on the same material demonstrate the presence of fish-tail effect.[34] Fish-tail effect can be considered to be one special form of peak effects. Among many origins for the peak effect, some well-know examples are due to softening of vortex lattice close to $H_{c2}$,[35] called synchronization, or geometrical matching of the vortex system to the underlying periodicity of the pinning potential in the sample.[36] In most cases, the peak in $J_c$ is rather sharp. In contrast to this, fish-tail effect in cuprate is, in general, defined by the broad peak in $J_c$-$H$ characteristics. The mechanism for this phenomenon can be classified into two, namely static and dynamic origins. In the former case, regions with lower $T_c$ in the sample turn into normal or close to normal, thus serving as pinning centers. Since the $T_c$ of such weak superconducting regions is expected to have broad distribution, the enhancement of $J_c$ occurs in a broad magnetic field range. Another mechanism is related to the dynamics of the vortex system. In the collective creep theory, creep exponent strongly depends on the size of the vortex bundle, which is a function of magnetic field.[37] In a single vortex regime, the creep exponent is µ=1/7 and it crossover to 3/2 in the small bundle regime. A smaller µ corresponds to faster motion of vortices, and hence the irreversible magnetization at the measuring time scale of the order of 1sec to 1min becomes small, resulting in an increase of *J* with magnetic field.[38] As the field is further increased, the creep exponent becomes 7/9 in the large bundle regime and the vortex creep becomes faster and the increase in $J_c$ with *H* would be weakened. However, the collective creep model is based on the elastic deformation of vortices. So, when the plastic deformations set in, the situation may change drastically. The elastic to plastic crossover scenario for the fish-tail effect is based on this kind of consideration. "crossover from elastic to plastic" is the most plausible origin of the fish-tail in the $Ca_{10}(Pt_{4-\delta}As_8)(Fe_{2-x}Pt_xAs_2)_5$ and other iron-based superconductors. It is partly supported by the strong acceleration of vortex motion above the peak field as demonstrated in various iron-based superconductors including $Ba(Fe,Co)_2As_2$.[25,39]

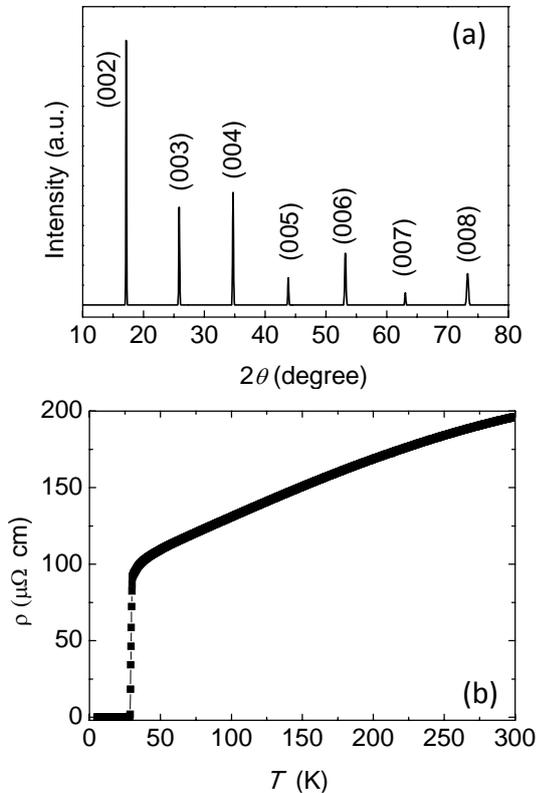

FIG. 1. (a) X-ray diffraction pattern of as-prepared single crystal. (b) Temperature dependence of in-plane resistivity of $Ca_{10}(Pt_{4-\delta}As_8)(Fe_{2-x}Pt_xAs_2)_5$ single crystal.

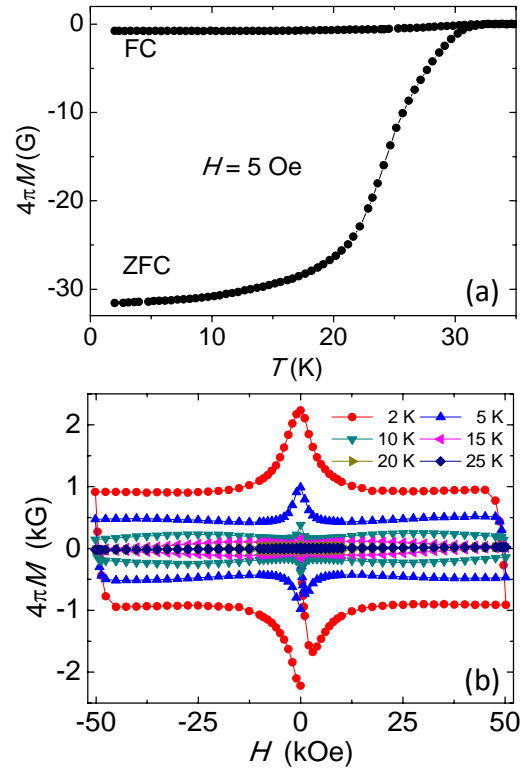

FiG. 2.(a) Temperature dependence of ZFC and FC magnetization in $Ca_{10}(Pt_{4-\delta}As_8)(Fe_{2-x}Pt_xAs_2)_5$ measured at 5 Oe. (b) Magnetic field dependence of magnetization in $Ca_{10}(Pt_{4-\delta}As_8)(Fe_{2-x}Pt_xAs_2)_5$ at different temperatures ranging from 2 to 25 K.

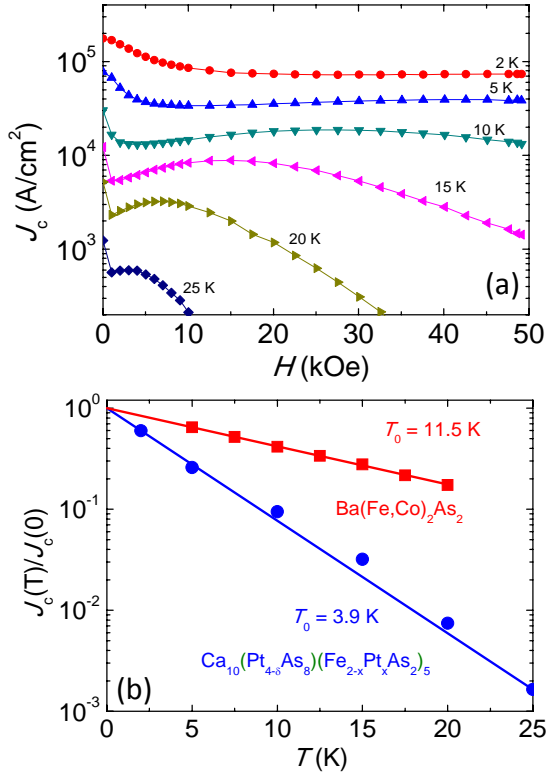

FIG. 3. (a) Magnetic field dependence of critical current densities obtained from the *M-H* curves shown in Fig. 2(b). (b) Critical current density $J_c$ relative to its value at absolute zero, dercreases roughly exponentially $\exp(-T/T_0)$, due to flux creep. With zero field, normalized $J_c$ for $Ca_{10}(Pt_{4-\delta}As_8)(Fe_{2-x}Pt_xAs_2)_5$ single crystal compared with that for $Ba(Fe,Co)_2As_2$ single crystals.

$J_c$ calculated from *M-H* curve is estimated to be $0.8\times10^5$ A/cm$^2$ at 5 K under zero field, which is smaller than that of optimally-doped $Ba(Fe_{1-x}Co_x)_2As_2$ single crystal, but similar to $FeTe_{0.6}Se_{0.4}$ single crystal,[40] and still in the range for applications. $J_c$ values in excess of $0.3\times10^5$ A/cm$^2$ are sustained up to 50 kOe at 5 K. One reason for the low $J_c$ is that since $Ca_{10}(Pt_{4-\delta}As_8)(Fe_{2-x}Pt_xAs_2)_5$ is anisotropic (discussed later), strong vortex dynamics suppresses *J* in the measured time window. In other words, $J_c$ can be larger at even lower temperatures, or if we measure in a much shorter time scale. Another possibility is that the material may have stacking fault due to the large Pt deficiency in Pt-As layer. In this case, $J_c$ would also be strongly suppressed.

Temperature dependence of reduced $J_c$ of $Ca_{10}(Pt_{4-\delta}As_8)(Fe_{2-x}Pt_xAs_2)_5$ under zero field is shown in Fig. 3(b). The temperature dependence of $J_c$ is well approximated by $J_c(T)=J_c(0)\exp(-T/T_0)$ with $T_0\approx3.9$ K, where $J_c(0)$ is the critical current density at zero temperature under zero field and $T_0$ is a characteristic temperature for vortex creep. The exponential temperature dependence arises from nonlinear effective flux-creep activation energy due to weak collective pinning of magnetic flux lines.[41-43] For comparison, temperature dependence of normalized $J_c$ of Ba(Fe$_{0.93}$Co$_{0.07}$)$_2$As$_2$ extracted from Ref. 25 is also plotted in this figure. As is evident, the Ca$_{10}$(Pt$_{4-\delta}$As$_8$)(Fe$_{2-x}$Pt$_x$As$_2$)$_5$ behaves like most widely studied iron-based superconductor Ba(Fe$_{1-x}$Co$_x$)$_2$As$_2$, but has stronger temperature dependence. This $J_c(T)$ behavior of Ca$_{10}$(Pt$_{4-\delta}$As$_8$)(Fe$_{2-x}$Pt$_x$As$_2$)$_5$ is also similar to that of YBa$_2$Cu$_3$O$_{7-\delta}$.[44,45] In general, $T_0$ is strongly correlated with the anisotropy of the superconducting materials. The smaller $T_0$ corresponds to larger anisotropy for superconductors with similar $T_c$. $T_0$ in Ba(Fe$_{0.93}$Co$_{0.07}$)$_2$As$_2$ is 11.5 K. The anisotropy of of Ca$_{10}$(Pt$_{4-\delta}$As$_8$)(Fe$_{2-x}$Pt$_x$As$_2$)$_5$ is much larger than that of Ba(Fe$_{0.93}$Co$_{0.07}$)$_2$As$_2$, but close to that of La$_{2-x}$Sr$_x$CuO$_{4-\delta}$,[46] indicating that the anisotropy in this superconductor is larger than that in Ba(Fe$_{0.93}$Co$_{0.07}$)$_2$As$_2$.

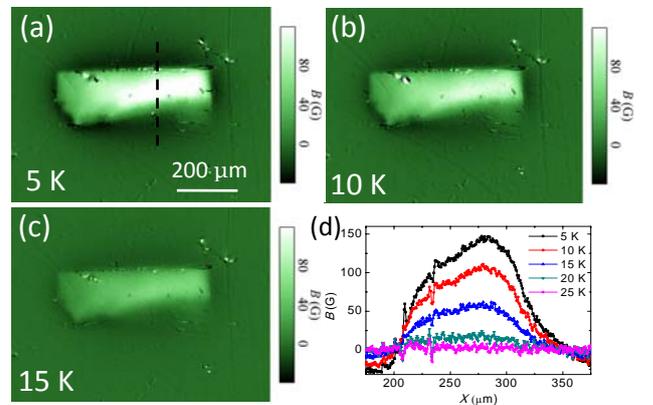

FIG. 4. (a) Magneto-optical images in the remanent state after applying 800 Oe in $Ca_{10}(Pt_{4-\delta}As_8)(Fe_{2-x}Pt_xAs_2)_5$ at 5, 10, and 15 K. (d) The local magnetic induction profiles at different temperatures taken along the dashed lines in (a).

The above estimate of $J_c$ relies on the assumption that homogeneous current is flowing within the sample. To examine this assumption, we made MO imaging of 19 μm thick Ca$_{10}$(Pt$_{4-\delta}$As$_8$)(Fe$_{2-x}$Pt$_x$As$_2$)$_5$ single crystal in the remanent state at several temperatures ranging from 5 to 25 K. The remanent state is prepared by applying 800 Oe along $c$ axis for 1 s and removing it after zero-field cooling. Figures 4(a)–(c) show MO images of Ca$_{10}$(Pt$_{4-\delta}$As$_8$)(Fe$_{2-x}$Pt$_x$As$_2$)$_5$ in the remanent state at 5, 10

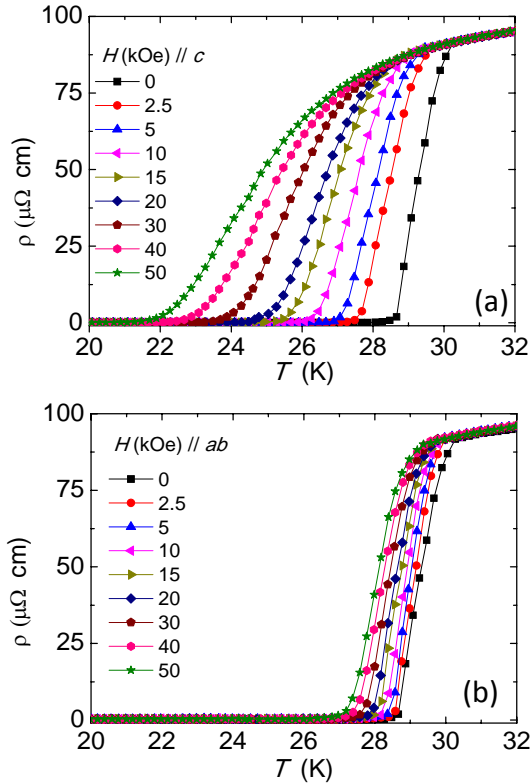

FIG. 5 Magnetic field dependence of in-plane resistivity for (a) $H \parallel c$ and (b) $H \parallel ab$.

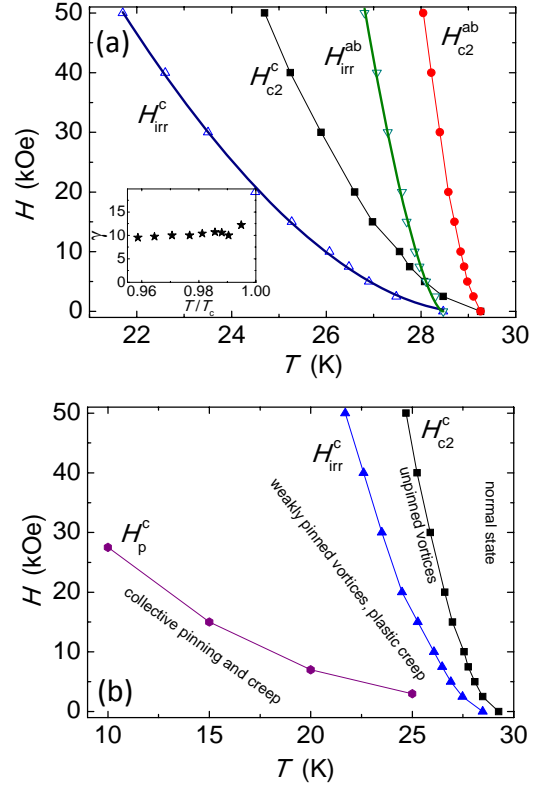

FIG. 6 (a) Temperature dependence of upper critical field and irreversible field for $H \parallel c$ and $H \parallel ab$ obtained by the 50% and 1% of normal state resistivity in Ca$_{10}$(Pt$_{4-\delta}$As$_8$)(Fe$_{2-x}$Pt$_x$As$_2$)$_5$ The solid lines are fits to $H_{irr}^{ab}(T) \approx (1-T/T_c)^{1.50}$ and $H_{irr}^{c}(T) \approx (1-T/T_c)^{1.75}$. The inset shows the the anisotropic parameter $\gamma(T) = H_{c2}^{ab}/H_{c2}^{c}$. (b) The $H$-$T$ phase diagram for $H \parallel c$ of Ca$_{10}$(Pt$_{4-\delta}$As$_8$)(Fe$_{2-x}$Pt$_x$As$_2$)$_5$ single crystals in the mixed state.

and 15 K, respectively. Figure 4(d) shows profiles of the magnetic induction at different temperatures along the line shown in Fig. 4(a). At lower temperatures, the MO image shows a nearly uniform current flow in the sample. $J_c$ for a thin superconductor is roughly estimated by $J_c \sim \Delta B/t$, where $\Delta B$ is the trapped field and $t$ is the thickness of the sample. With $\Delta B \sim 150$ G and $t=19$ μm, $J_c$ is estimated as $\sim 0.8\times 10^5$ A/cm$^2$ at 5 K, which is in good agreement with the value obtained from $M$-$H$ curve.

The variations of $T_c$ with magnetic field from 0 to 50 kOe for $H \parallel c$ and $H \parallel ab$ are shown in Fig. 5 (a) and (b), respectively. With increasing field, the resistive transition shifts to lower temperatures accompanied by an increase in the transition width, especially for $H \parallel c$, which also indicates the presence of strong thermal fluctuation of the vortices. The broadening is almost negligible for $H \parallel ab$. With 50 kOe field, the $T_c$ was suppressed to $0.76T_{c0}$ for $H \parallel c$ and $0.94T_{c0}$ for $H \parallel ab$, where $T_{c0}$ is the transition temperature under zero field.

Figure 6 (a) shows the variation of the upper critical field $H_{c2}$ and irreversibility field $H_{irr}$ with temperature for Ca$_{10}$(Pt$_{4-\delta}$As$_8$)(Fe$_{2-x}$Pt$_x$As$_2$)$_5$. The values of $H_{c2}$ are defined as the field at the midpoint of the resistive transition. The slopes of $H_{c2}$ are -16.8 kOe/K and -57.1 kOe/K along $c$ and $ab$ directions, respectively, extracted from the linear part between 30 kOe and 50 kOe. From the Werthamer–Helfand–Hohenberg theory,[47] which describes orbital depairing field for conventional dirty type II superconductors, the value of $H_{c2}$ at $T=0$ K is estimated using $H_{c2}(0)=0.69T_c|dH_{c2}/dT|_{T=Tc}$. $H_{c2}^c(0)$ and $H_{c2}^{ab}(0)$ are estimated as 339 kOe and 1154 kOe, respectively. The anisotropy parameter $\gamma = H_{c2}^{ab}/H_{c2}^c = \sqrt{m_c^*/m_{ab}^*}$ shown in inset is around 10 near $T_{c0}$. The anisotropy of $Ca_{10}(Pt_{4-\delta}As_8)(Fe_{2-x}Pt_xAs_2)_5$ is much larger than the typical value around 2~3 in $Ba(Fe_{0.93}Co_{0.07})_2As_2$.[25] This is consistent with the tendency inferred from the temperature dependence of the critical current density measurement. The anisotropy of our sample is also larger than that in Ref. 21, the higher $T_c$ of our sample may be the reason for this difference.

The large anisotropy together with the broadening of the superconducting transition under magnetic field suggests the existence of a wide vortex-liquid phase when the magnetic field is applied along the c-axis. $Ca_{10}(Pt_{4-\delta}As_8)(Fe_{2-x}Pt_xAs_2)_5$ crystal is not the only superconductor in iron-based superconductors having relatively large anisotropy. For example, 1111 shows also large anisotropy.[48] There have been several reports showing the presence of reversible region, which corresponds to the vortex liquid phase in global sense.[48-50] In this sense, the separation of $H_{irr}$ and $H_{c2}$ is the evidence for the presence of vortex liquid phase. However, no vortex solid to liquid transitions accompanying clear first-order signature of transition have been reported.[51,52]

In order to extract the superconducting parameters, we have used the Ginzburg–Landau (GL) formula for the coherence length ($\xi$). $\xi$ is calculated from the estimated $H_{c2}^{ab}(0)$'s data using the relations given by $H_{c2}^{ab}(0) = \Phi_0/2\pi\xi_{ab}\xi_c$, $H_{c2}^c(0) = \Phi_0/2\pi\xi_{ab}^2$, where $\Phi_0=2.07\times10^{-7}$ G cm$^2$, the obtained $\xi_{ab}$ ~1.74 nm and $\xi_c$~5.92 nm. The values of $H_{irr}$ were defined as the field at 1% of the normal state resistivity. The temperature dependence of the irreversibility field is well approximate by $H_{irr}^{ab}(T) \approx (1-T/T_c)^{1.50}$ and $H_{irr}^c(T) \approx (1-T/T_c)^{1.75}$. Such $(1-T/T_c)^n$ type temperature dependence of $H_{irr}$ can be explained by using a thermally activated flux creep model.[37] The power exponent 1.50 for $H_{irr}^{ab}(T)$ of $Ca_{10}(Pt_{4-\delta}As_8)(Fe_{2-x}Pt_xAs_2)_5$ is close to that of $YBa_2Cu_3O_{7-\delta}$[53] and $Ba(Fe_{0.93}Co_{0.07})_2As_2$.[54]

Adopting similar analyses in Ref. 30, the phase diagram of $Ca_{10}(Pt_{4-\delta}As_8)(Fe_{2-x}Pt_xAs_2)_5$ single crystals in the mixed state is plotted in Fig. 6 (b). Here $H_p$ is the magnetic field of the peak position in M-H curve. This phase diagram is also similar to that of $Ba(Fe_{1-x}Co_x)_2As_2$,[39,54] $Ba_{1-x}K_xFe_2As_2$[55] and high-$T_c$ cuprates.

## IV. CONCLUSION

In conclusion, X-ray diffraction, magnetization, resistivity, and magneto-optical measurements were performed on high quality $Ca_{10}(Pt_{4-\delta}As_8)(Fe_{2-x}Pt_xAs_2)_5$ single crystals. Magneto-optical imaging revealed nearly homogeneous current flow in the crystal. The magnetization measurements reveal fish-tail hysteresis loop in the intermediate temperature range. The value of $J_c$ is over $0.8\times10^5$ A/cm$^2$ below 5 K under zero field, which is promising for practical applications. Upper critical fields obtained by resistive transition are 339 kOe and 1154 kOe at zero temperature along c and ab directions, respectively. The anisotropy parameter $\gamma$ near $T_c$ is around 10. The temperature dependence of $J_c$ also suggests that the anisotropy in $Ca_{10}(Pt_{4-\delta}As_8)(Fe_{2-x}Pt_xAs_2)_5$ is much larger than that of $Ba(Fe_{0.93}Co_{0.07})_2As_2$.